\documentclass[a4paper]{article}

\usepackage{INTERSPEECH2021}
\newcommand{\argmax}{\operatornamewithlimits{argmax}}

\usepackage{caption}
\usepackage{subcaption}
\usepackage{amssymb}
\usepackage{multirow}
\usepackage{float}
\usepackage{lipsum}
\usepackage{hyperref}

\usepackage[capitalize]{cleveref}
\Crefname{chapter}{Chap.}{Chaps.}
\Crefname{section}{Sec.}{Secs.}
\Crefname{figure}{Fig.}{Figs.}
\Crefname{table}{Table}{Tables.}

\usepackage{pifont}

\usepackage{array}
\newcolumntype{?}{!{\vrule width 1.5pt}}

\usepackage{enumerate} 

\title{Towards Consistent Hybrid HMM Acoustic Modeling}
\name{Tina Raissi$^1$, Eugen Beck$^{2}$, Ralf Schl\"uter$^{1,2}$, Hermann Ney$^{1,2}$}
\address{
	$^1$Human Language Technology and Pattern Recognition Group, RWTH Aachen University, Germany \\ $^2$AppTek GmbH, Aachen, Germany}
\email{\{raissi,beck,schlueter,ney\}@cs.rwth-aachen.de}

\begin{document}

\maketitle
\begin{abstract}
High-performance hybrid automatic speech recognition~(ASR) systems are often trained with clustered triphone outputs, and thus require a complex training pipeline to generate the clustering.\ The same complex pipeline is often utilized in order to generate an alignment for use in frame-wise cross-entropy training.\ In this work, we propose a flat-start factored hybrid model trained by modeling the full set of triphone states explicitly without relying on clustering methods.\ This greatly simplifies the training of new models.\ Furthermore, we study the effect of different alignments used for Viterbi training.\ Our proposed models achieve competitive performance on the Switchboard task compared to systems using clustered triphones and other flat-start models in the literature.

\end{abstract}
\vspace{-0.2cm}
\section{Introduction}
One of the early applications of neural networks~(NNs) in automatic speech recognition~(ASR) task was within the hidden Markov model~(HMM) framework~\cite{bourlard2012connectionist}.\ By defining a set of states which cannot be observed, the HMM establishes a relation between a given sequence of features or observations and the sequence of output symbols.\ The likelihood of generating a feature observation given a state, originally estimated by a Gaussian mixture model~(GMM), is replaced in the hybrid approach by normalized scaled state posteriors estimated by a neural model.\ The HMM labels for the states are context-dependent units, such as triphones, i.~e., phonemes augmented with their left and right phonetic contexts.\ Due to the large number of triphone state labels and for a robust parameter estimation, the states are usually clustered with classification and regression trees~(CART)~\cite{young1994tree}.\ The Viterbi training of the hybrid model also requires an initial alignment of the tied-states to the acoustic features, typically obtained from a previously trained GMM.\ Moreover, the overall system in the standard configuration makes use of a pronunciation lexicon, a phoneme inventory, and a language model.\ The aforementioned components of the hybrid system require several training stages and do not allow for a consistent and unified acoustic modeling.\ The end-to-end~(E2E) encoder-decoder models~\cite{ctc, rnnt, rnaSak,attention} on the other hand offer the possibility of carrying out a joint optimization of different components, by directly transforming a sequence of feature vectors representing the speech signal, to a sequence of characters or wordpieces.\ Despite the recent advances in E2E modeling~\cite{wei, zoltan, tsukernnt}, hybrid models continue to be competitive~\cite{facebook,transformer}.\ Moreover, in the speech community, their advantage is often underlined when low-resource tasks are involved.\ Recently it was also shown that for the domain adaptation tasks with acoustic or content mismatch the hybrid approach can lead to larger improvements, so far~\cite{pawel}.

In our previous work\cite{raissi}, we proposed the factored hybrid model: a context-dependent~(CD) HMM/NN hybrid model that does not use phone clustering for the computation of label posteriors.\ We showed that given an existing alignment from a tandem system\cite{tuske}, it is possible to attain similar performance as a CART based system.\ In this work, we take a further step towards the possible elimination of the system dependency on an external alignment.\ Specifically, we do a two-fold investigation: (1)  flat-start\footnote{We borrow this terminology from \cite{povey}, and use it for denoting a GMM-free hybrid model built from scratch.} training of a monophone factored hybrid, (2) Multi-stage phonetic training of our proposed triphone with fixed path and frame-wise cross-entropy~(CE), starting with different alignments.\ There are multiple early works in the literature which propose hybrid models trained with the full-sum criterion, i.\ e.\ without the Viterbi approximation\cite{bw1,bw2,bw3}.\ In these works it has been argued that the soft alignment on the frame level is more useful for ASR, because of the inherently fuzzy nature of phoneme boundaries and the overlap in the windows used for the feature extraction.\ Today, with bidirectional Long Short-Term Memory~(Bi-LSTM) based encoders, as argued in \cite{zeyerbeckctc}, the scenario has slightly changed.\ There is an additional issue with one label dominating the learned alignments due to the access of the encoder to the full sequence.\ Moreover, the training of a network which starts from scratch can be unstable.\ So far, with the exception of the work using Time-Delay NN\cite{endtoendpovey}, none of the prior works considered the unclustered set of states for full-sum training of hybrid model.\ In addition to the proposed flat-start factored hybrid, the objective of our current work is also to study the behavior of our models with respect to the choice of different alignments in CE training with Viterbi.\ It is important to note that in this paper we apply a simplifying assumption in our full-sum training objective, as described in \cref{subsec:tc}.\ In the next sections we will present our framework and the models, followed by experimental results and our conclusions.
\vspace{-0.1cm}
\section{Modeling Framework and Training}
\label{sec:modelapproach}

\subsection{Standard Hybrid Approach}
\label{subsub:standard}
\vspace{-0.1cm}
The statistical formulation of the ASR task maximizes the a-posteriori probability of the word sequence $w_1^N$ given the input features $x_1^T$, with $T \gg N$, based on the \textit{Bayes decision rule}~\cite{bayes1763lii}, as follows:
\vspace{-0.2cm}
\begin{subequations}	
\begin{align}
x_1^T \rightarrow\tilde{w}_1^{\tilde{N}}(x_1^T) = & \underset{N, w_1^N}{\argmax} \left\lbrace p(w_1^N|x_1^T) \right\rbrace \notag\\ 
= & \underset{N, w_1^N}{\argmax} \left\lbrace p(x_1^T|w_1^N) \cdot p(w_1^N) \right\rbrace  \label{eq:bayes2}
\end{align}
\end{subequations}
The acoustic component $p(x_1^T|w_1^N)$ in \cref{eq:bayes2} modeling the probability of observing this feature sequence given a word sequence, in the HMM framework can be rewritten as follows:\vspace{-0.2cm}
\begin{equation}\footnotesize 
 \hspace{-0.2cm}p(x_1^T|w_1^N) = \underset{ s_1^T}{\text{max}} \hspace{-0.05cm} \prod_{t=1}^{T} p(x_t | s_t, w_1^N; \phi_1^M) p(s_t |  s_{t-1}, w_1^N;\phi_1^M)\label{eq:vitdecode}
\vspace{-0.05cm}
\end{equation}
where after summing over all possible hidden state sequences, we apply Markov and output independence assumptions and factorize into emission and state transition probabilities.\ As a further model assumption we substitute the summation with maximization, for a Viterbi decoding~\cite{viterbi}.\ Moreover, $\phi_1^M$ is the triphone sequence of length $M$ corresponding to the word sequence.\ For simplicity, here the pronunciation distribution can be assumed to be deterministic.

\subsection{Direct Modeling of Context}
\label{sub:ctx}

In the standard hybrid model $s$ corresponds to the label of the respective cluster of tied-states.\ In our case, we do not tie the labels.\ With similar notation to~\cite{raissi}, denote by $\left\lbrace \phi_{\ell},\phi_{c},\phi_{r}\right\rbrace_t$ the set of left, center and right phonemes of the aligned triphone at time frame $t$.\ We model the triphone state at time frame $t$ explicitly by means of a state class $c(s_t, w_1^N) = \left\lbrace \phi_{\ell},\phi_{c},\phi_{r}, i \right\rbrace _t$, where $i$ enumerates the HMM state of the corresponding triphone.\ Define $p_{_t}$ to be the parametrized probability distribution at time frame $t$.\ The emission probability of \cref{eq:vitdecode} can now be reformulated as follows:
\begin{equation}
\label{eq:emis}\nonumber
\hspace{-0.3cm}p(x_t|s_t, \phi_1^M, w_1^N)=p(x_t | c(s_t, w_1^N)) =p_{_t}(x| \phi_{\ell},\phi_{c},\phi_{r}, i) \hspace{-0.1cm}
\end{equation}
For simplicity, we define $\sigma_c = (\phi_c, i)$ to be the center phoneme state identity.\ For notational simplicity, the likelihood of observing a feature vector $x_t$ is consequently written as follows:
$$\label{eq:sigma_c}
p(x_t | \phi_t^{\ell},\phi_t^{c},\phi_t^{r}, i_t) = p_{_t}(x | \phi_{\ell},\phi_{c},\phi_{r}, i) = p_t(x|\phi_{\ell},\sigma_c,\phi_{r}) $$
By using the usual substitution in the hybrid approach via Bayes identity, it is possible to replace $p(x|\phi_{\ell},\sigma_c,\phi_{r})$ by the locally normalized joint probability of \cref{eq:normjoint}, with class-conditional acoustic model~(AM) and prior scales, $\alpha$ and $\beta$.\ \vspace{-0.2cm}
 \begin{subequations}
 \begin{align}
 p_t(x | \phi_{\ell},\sigma_{c},\phi_{r})&=\frac{p_t(\phi_{\ell},\sigma_{c},\phi_{r} | x)\cdot p_t(x)}{p(\phi_{\ell},\sigma_{c},\phi_{r})} \nonumber\\ 
  &\sim \frac{p_t(\phi_{\ell},\sigma_{c},\phi_{r} | x)^{\alpha}}{p(\phi_{\ell},\sigma_{c},\phi_{r})^{\beta}} \label{eq:normjoint} 
 \end{align}
 \end{subequations}
The resulting state labels refer to the simple enumeration of all possible triphone states, where the identity of each state is uniquely defined by its position within the center phoneme and its right and left contexts.\ Given that in our case each phoneme consists of three HMM states, we would need $n=3\times |phonemes|^3$ softmax outputs, e.g., around 300k for the Switchboard task, for the joint posterior in numerator of \cref{eq:normjoint}.\ Different early works on CD hybrid models tried to address this issue by using factorized neural networks~\cite{bourlard1992cdnn, franco1994context}.\ However, the joint posterior probability modeled by the mentioned works still takes into account the set of clustered states.\
Depending on the type of context we want to model, it is possible to carry out different factorizations.\ We consider two main CD models and a monophone model as described in the following paragraphs.\ The monophone model is relevant in our work for two reasons: (1) a multi-stage phonetic training with regularization effect on the model performance, (2) an initial study presented in this work on full-sum training of a hybrid model without state-tying. \vspace{-0.1cm}
\vspace{-0.1cm}
\subsubsection{Decision Rules}
\vspace{-0.1cm}
\label{dr}
The decoding of our CD and monophone models relies on the application of \cref{eq:vitdecode} via the suitable factorization of the joint posterior obtained in \cref{eq:normjoint}.\ Each factor refers to a softmax output belonging to the context-dependent neural network trained in a multi-task manner, either with full-sum or with the fixed best path.\ The architectures of different models are depicted in \cref{fig:archs}.\ Each decision rule consists of a subset of the trained outputs and is defined following the mathematically sound factorization and definition of dependencies.\ This means, the trained model can have additional outputs that are not used during decoding.\ Moreover, we set AM scale $\alpha$ to one and define separate prior scales for each CD prior in the denominator.
Among different possible decompositions, for our triphone model we consider a left-to-right trigram, via a forward decomposition.\ \vspace{-0.3cm}
\begin{subequations}
	\label{eq:tri-fwd}
	\begin{align}
	\hspace{-0.33cm} p_t(x | \phi_{\ell},\sigma_c,\phi_{r}) &= \frac{p_t(\phi_{\ell},\sigma_c, \phi_{r} |x) p_t(x)}{p(\phi_{\ell}, \sigma_c, \phi_{r})} \nonumber \\
	&\sim \frac{p_t(\phi_{r}|\phi_{\ell},\sigma_c, x)p_t(\sigma_c|\phi_{\ell}, x) p_t(\phi_{\ell}|x)}{p(\phi_{r} | \phi_{\ell},\sigma_c)^{\beta^{\prime \prime}}p(\sigma_c|\phi_{\ell})^{\beta^{\prime}} p(\phi_{\ell})^{\beta}} 
	\end{align}	
\end{subequations} 
For the monophone models presented in this work we only use the center state phoneme, as shown in \cref{eq:mono}. Moreover, we add the dependency to the left context for diphone models, as described in \cref{eq:di}. \vspace{-0.3cm}
\begin{equation}
\label{eq:mono}
p_t(x | \sigma_c) \sim \frac{p_t(\sigma_c|x)}{p(\sigma_c)^{\beta}}
\end{equation}
\vspace{-0.6cm}
\begin{equation}
 	\label{eq:di}
 	p_t(x | \sigma_c,\phi_{\ell}) \sim \frac{p_t(\sigma_c|\phi_{\ell}, x) p_t(\phi_{\ell}|x)}{p(\sigma_c|\phi_{\ell})^{\beta^\prime} p(\phi_{\ell})^{\beta}} 	
\end{equation}
\vspace{-0.3cm}
\subsection{Training Criteria}
 \vspace{-0.1cm}
\label{subsec:tc}
 \subsubsection{Flat-Starting}
 \vspace{-0.2cm}
The maximum log likelihood criterion for training the acoustic model of \cref{eq:bayes2}, modeled with the set of parameters $\theta$, is defined as:\vspace{-0.2cm}
\begin{equation}
\label{ml} \vspace{-0.2cm} 
\underset{\theta}{\text{max }} \left\lbrace \log L(\theta) \right\rbrace := \underset{\theta}{\text{max }} \left\lbrace \log p(x_1^T|w_1^N, \theta) \right\rbrace  
\end{equation}
We calculate the derivative similarly to \cite{zeyerbeckctc}.\ We choose as our starting point the following formulation used in the standard hybrid approach, with $c(s)$ and $\gamma(c(s)|x_1^T, w_1^N)$ being the CART label and its posterior occupancy, respectively:\vspace{-0.1cm}
\begin{equation}
\footnotesize \vspace{-0.1cm}
\frac{\partial}{\partial \theta}\log L(\theta)=\sum_{t,c(s)}\gamma_t(c(s)|x_1^T,w_1^N, \theta) \cdot \frac{\partial}{\partial \theta}\log p_{_t}(x|c(s), \theta) \nonumber
\end{equation}
 By limiting the summation over the set of valid state sequences corresponding to the set of word sequences, we have:\vspace{-0.1cm}
\begin{equation}
\label{eq:gamma} \footnotesize  \nonumber \vspace{-0.1cm}
\gamma_t(c(s)|x_1^T,w_1^N, \theta) = \frac{\sum_{\{s_1^T:w_1^N,s_t =c(s)\}} p(x_1^T, s_1^T | w_1^N, \theta)}{\sum_{s_1^T:w_1^N} \hfill p(x_1^T, s_1^T | w_1^N, \theta)}
\end{equation}
In our proposed approach, we plug in our definition of a state class, as described in \cref{sub:ctx} and carry out the factorization for a triphone model with forward decomposition.\ This leads to separate factors for the log probabilities, each weighted by the state class occupancy.\ The latter can then be marginalized over all random variables that differ from the one occurring in the log probability.\ This will lead to three different gamma values $\gamma(z|\cdot), \text{ for } z \in \{\phi_{\ell}, \sigma, \phi_r\}$, that we denote as left context, center state and right context occupancy, respectively.\ Moreover, as an initial study case, we apply a simplifying assumption for the log probabilities by considering the dependency only on the input signal and not the other phonetic or phoneme state entities, as shown in \cref{eq:simple}.\ The training procedure consequently is similar to a generalized Expectation-Maximization algorithm, where we iteratively alternate the estimation of the contexts and center state occupancies with the NN backpropagation step.\ \vspace{-0.1cm}

\begin{subequations}
	\footnotesize
\begin{align}
\nonumber 
  \vspace{2cm}\hspace{-0.2cm}\frac{\partial}{\partial \theta}\log L(\theta)=&\hspace{-0.5cm}\sum_{t,(\phi_{\ell}, \sigma_c, \phi_r)}\hspace{-0.5cm}\gamma_t((\phi_{\ell}, \sigma_c, \phi_r)|x_1^T,w_1^N, \theta)  \cdot \frac{\partial}{\partial \theta} \log p_{_t}(x|\phi_{\ell}, \sigma_c, \phi_r, \theta)\\
 \label{eq:complete}
 \sim \hspace{-0.5cm}\sum_{t, \phi_{\ell}, \sigma_c, \phi_r}\hspace{-0.3cm}&\gamma_t((\phi_{\ell}, \sigma_c, \phi_r)|x_1^T,w_1^N, \theta)  \\  
 \nonumber   
 \cdot \frac{\partial}{\partial \theta}\log &\left( \frac{p_t(\phi_{r}|\phi_{\ell},\sigma_c, x, \theta)p_t(\sigma_c|\phi_{\ell}, x, \theta) p_t(\phi_{\ell}|x, \theta)}{p(\phi_{r} | \phi_{\ell},\sigma_c)^{\beta^{\prime \prime}}p(\sigma_c|\phi_{\ell})^{\beta^{\prime}} p(\phi_{\ell})^{\beta}}\right) \\ 
\nonumber 
 =^{^{\hspace{-1cm}\substack{ \text{simplifying} \\ \text{assumption}}}}\sum_{t}&\left( \sum_{\phi_{\ell}} \gamma_t(\phi_{\ell}|x_1^T,w_1^N, \theta)\cdot \frac{\partial}{\partial \theta}\log \frac{p_t(\phi_{\ell}|x,\theta) }{p(\phi_{\ell})^{\beta^{\prime \prime}}} \right.\\
\nonumber
&\left.+\sum_{\sigma_c}\gamma_t(\sigma_c|x_1^T,w_1^N, \theta)\cdot\frac{\partial}{\partial \theta} \log \frac{p_t(\sigma_c| x,\theta) }{p(\sigma_c)^{\beta^{\prime}}}\right.\\
&\left .+\sum_{\phi_{r}}\gamma_t(\phi_{r}|x_1^T,w_1^N, \theta)\cdot \frac{\partial}{\partial \theta}\log \frac{ p_t(\phi_{r}|x, \theta)}{p(\phi_{r})^{\beta}}\right)\hspace{-0.1cm} \hspace{-1cm}\label{eq:simple}
\end{align}
\end{subequations}
 It is possible to approximate the sum over the hidden state sequences in \cref{eq:complete} with the maximum and use Viterbi.\ Instead of maximizing the likelihood of the observed data, Viterbi training maximizes the probability of the most likely state sequence.\ In this case, $\gamma_t(c(s))$ is one-hot encoding of the state class aligned in the best path.\ The best path calculated with Viterbi approximation can be kept fixed for all epochs and can be derived from external sources.\ The model is trained using frame-wise cross-entropy.\ We compare the Viterbi training with fixed path starting with alignments from both GMM based systems and our proposed forced-aligned flat-start CD model.\ Moreover, The CD models using Viterbi and unclustered set of states use pre-training on phonetic level\cite{raissi}.\
 \begin{figure}[t]
 	\centering
 	\vspace{-0.4cm}
 	\subcaptionbox{\scriptsize  Standard hybrid with CART.\label{fig:monocart}}{		
 		\includegraphics[height=0.30\columnwidth, trim={11cm 6cm 11cm 6cm},clip]{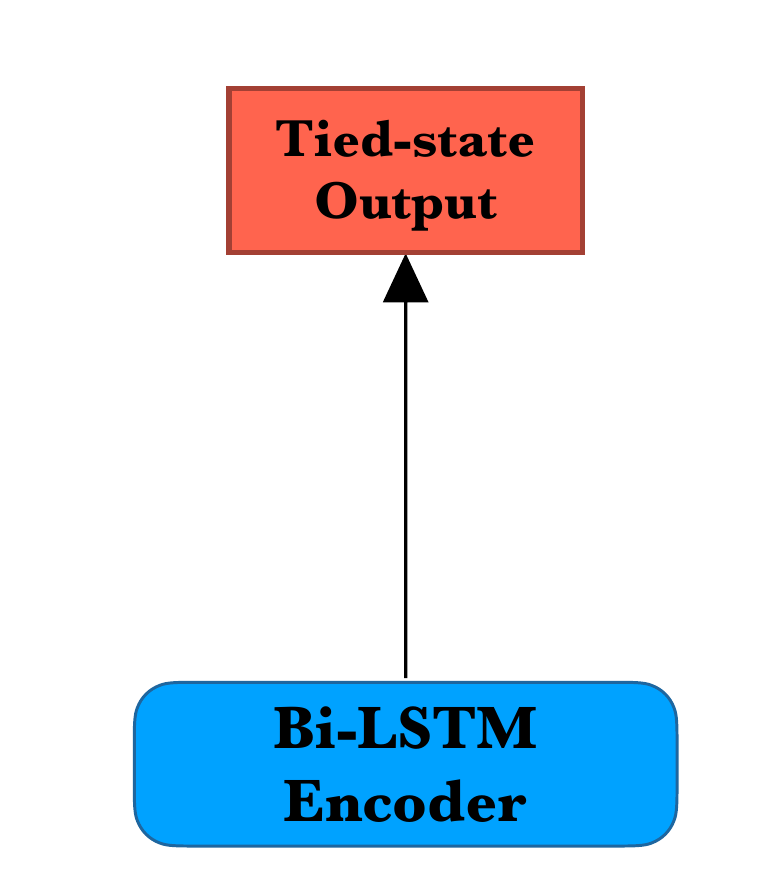}}
 	\quad 
 	\subcaptionbox{\scriptsize Factored hybrid (simplified).\label{fig:monofact}}{
 		\includegraphics[height=0.30\columnwidth, trim={8cm 6cm 8cm 6cm},clip]{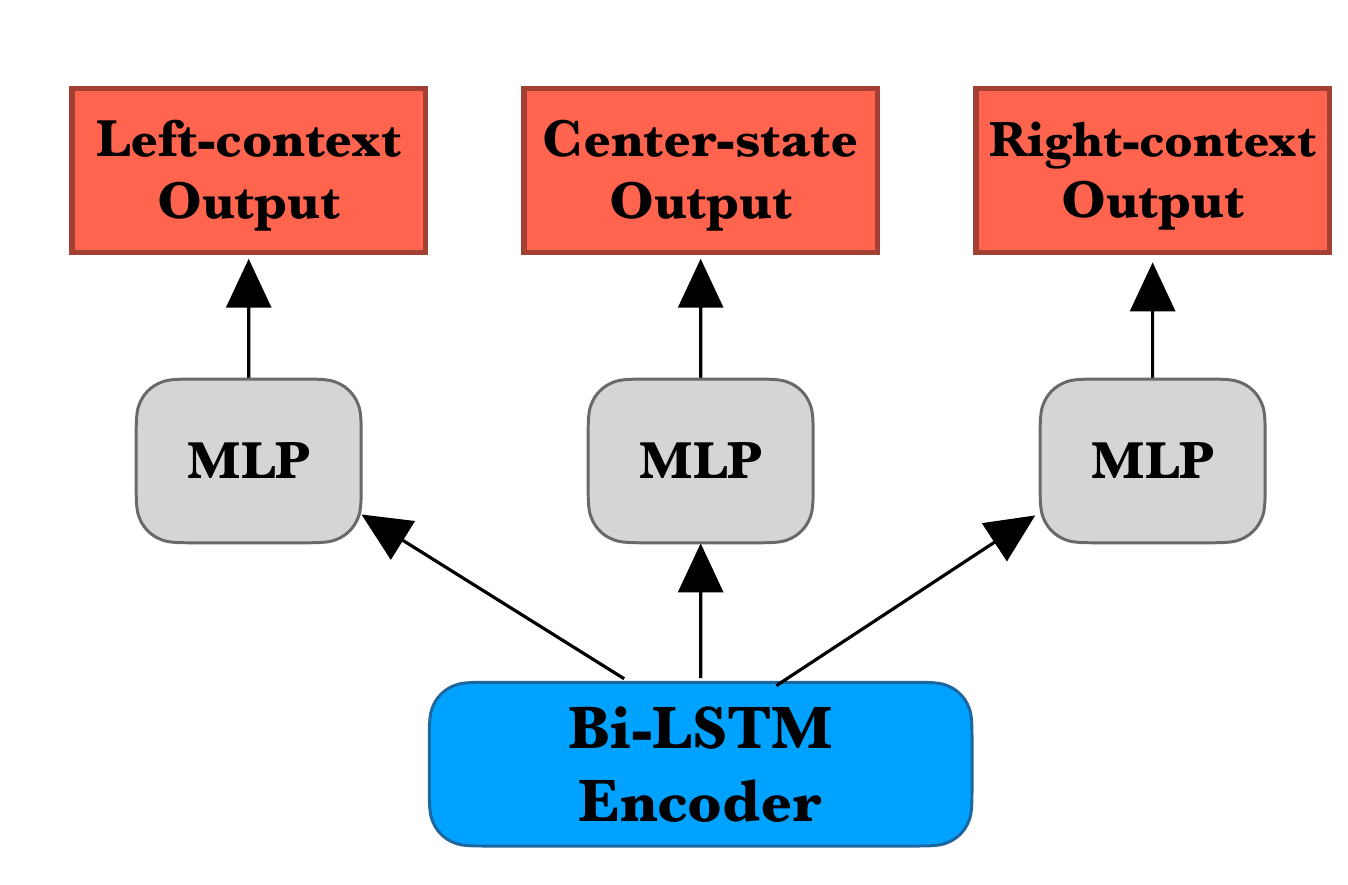}}
 	\subcaptionbox{\scriptsize Factored hybrid with Baum-Welch (simplified).\label{fig:bw}}{
 		\includegraphics[height=0.57\columnwidth, trim={5cm 3cm 5cm 3cm},clip]{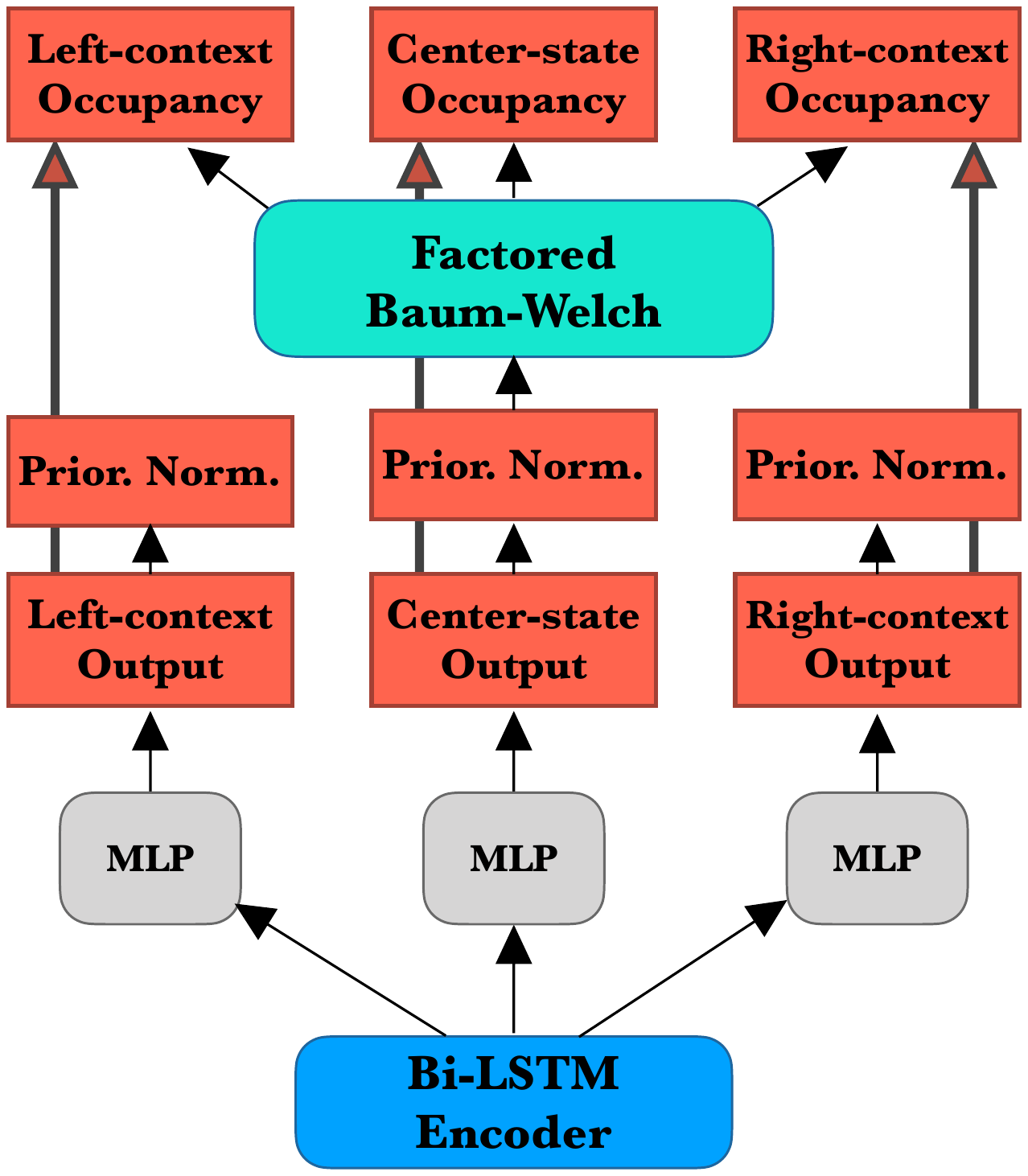}
 	}
 	\subcaptionbox{\scriptsize Diphone factored hybrid.\label{fig:di}}{		
 		\includegraphics[height=0.32\columnwidth, trim={8.1cm 6cm 8.1cm 6cm},clip]{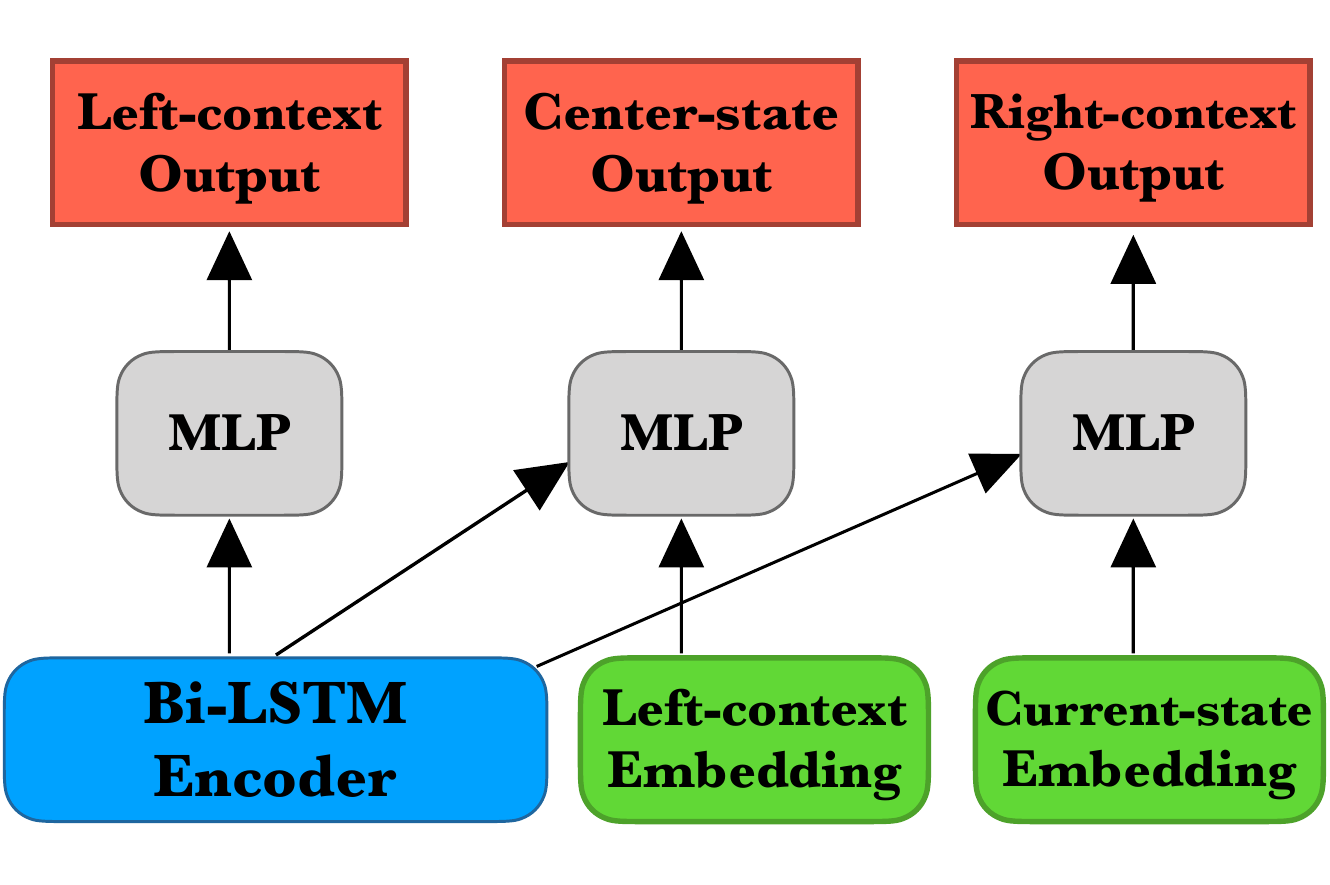}}
 	\hspace{-0.2cm}
 	\subcaptionbox{\scriptsize Triphone factored hybrid.\label{fig:fwd}}{
 		\includegraphics[height=0.32\columnwidth, trim={8.1cm 6cm 8.1cm 6cm},clip]{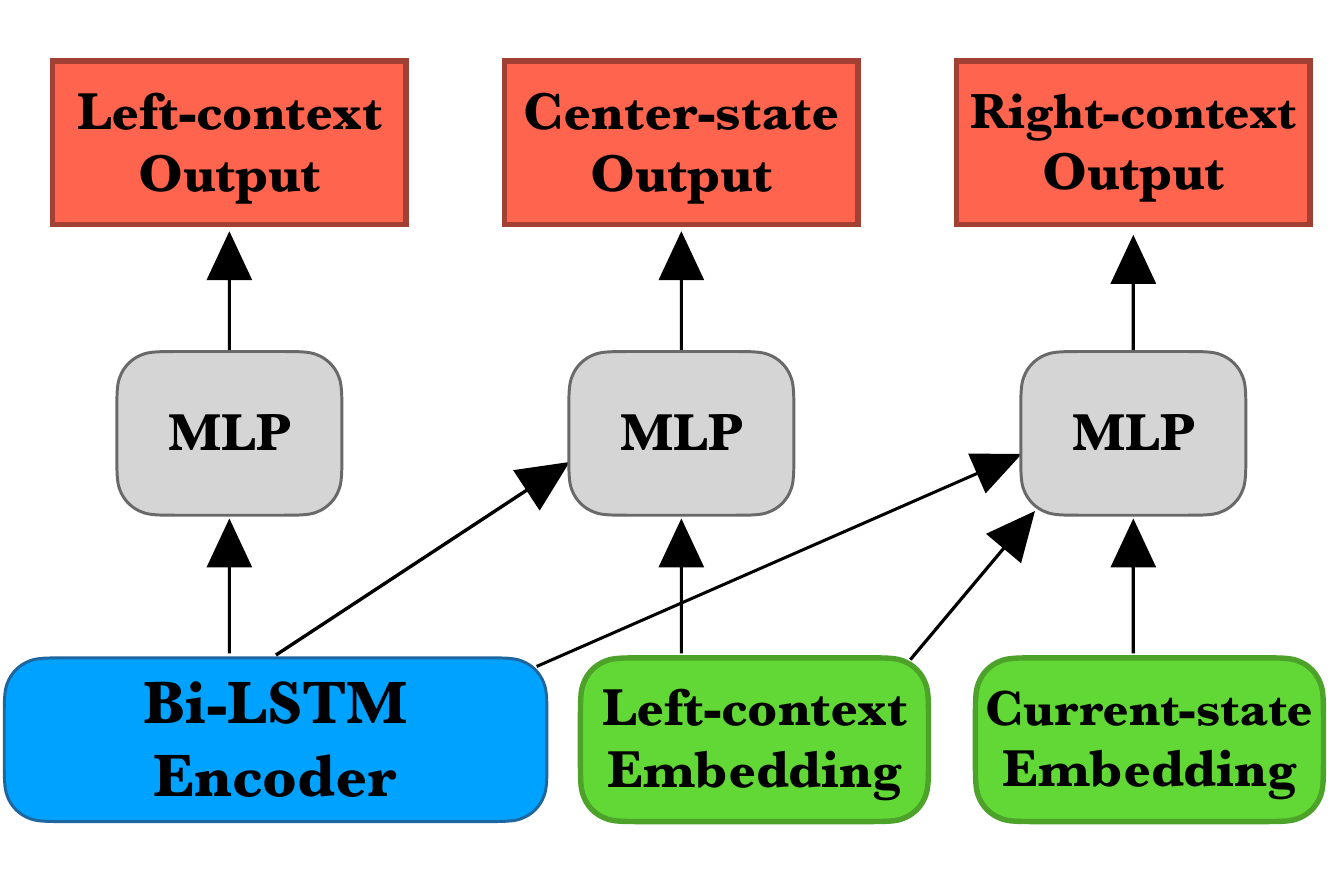}}
 	\vspace{-0.2cm} 	
 	\caption{Comparison of the architecture of different models with and without phone clustering.\ The model in \cref{fig:monocart} uses CART labels.\ With unclustered states and factorization we have the architecture depicted in \cref{fig:monofact}, where we drop the dependencies to the context for each factor.\ Both models use Viterbi.\ The simplified factored hybrid trained with full-sum, is shown in \cref{fig:bw}.\ The two architectures of \cref{fig:di,fig:fwd} are used for providing the posterior probabilities of \cref{eq:tri-fwd,eq:di}.\ }
 	\label{fig:archs} 	\vspace{-0.7cm}
 \end{figure}
 
   \begin{table*}[t]
   	\setlength{\tabcolsep}{0.3em}\renewcommand{\arraystretch}{1.03} 
   	\centering \footnotesize
   	   	\caption{An overview of alignment models used for Viterbi training of our factored hybrid models, depicted in \cref{fig:archs}.\ All results are on {\normalfont Hub'00} using 4-gram LM.\ For each alignment, number of stages, and if applicable, epochs~(Eps.), together with the architecture~(Arch.) and the performance of the model on {\normalfont HUB'00} are reported.\ The context-dependency both during the stages and for the final model are denoted by Mono, Di and Tri for Mono-, Di-, and Triphone, respectively.\ In \cref{tab:mono} we report WERs using the alignments from the models described in \cref{tab:align}.\ }
   	\begin{subtable}{.7\textwidth}
   		\centering \vspace{-0.25cm}
   		\caption{Different alignment models used for CE training with Viterbi.}
   		\label{tab:align}
   		\vspace{-0.2cm}	
   		\begin{tabular}{|c|c|c|c|c|c|c|} 
   			\hline		
   			\textbf{Name} &\textbf{Model} & \textbf{Arch.}&\textbf{Context} &\textbf{\#Stages} &\textbf{\#Eps} & \textbf{WER[\%]} \\ \hline
   			\textbf{Tandem}  &\cite{tuske}&\multirow{2}{*}{n/a }& Tri& 7& n/a & $13.9$ \\  \cline{1-2}\cline{4-7}
   			\textbf{GMM-Mono} &GMM&&\multirow{2}{*}{Mono}& 1& n/a& $71.3$\\  \cline{1-3}\cline{5-7}
   			\textbf{FH-Mono}&\multirow{2}{*}{FH}&\cref{fig:monofact}&&2&30&$17.1$\\ \cline{1-1}\cline{3-7}
   			\textbf{FH-Di}&&\cref{fig:di}&Di& 3&57&$15.4$\\ \hline
   			\textbf{FS-FH-Mono}&Flat-start FH&\cref{fig:bw}&Mono&0&30& $20.1$\\ \hline		
   		\end{tabular}  	
   		
   	\end{subtable}
   	\begin{subtable}{.3\textwidth}
   		\centering \vspace{-0.6cm}
   		\caption{Different monophone models trained with alignments from \cref{tab:align}.\ }
   		\label{tab:mono}
   		\vspace{-0.2cm}	
   		\begin{tabular}{|c|c|c|} 
   			\hline
   			\textbf{Model} & \textbf{Alignment}&\textbf{WER[\%]} \\ \hline
   			\multirow{4}{*}{FH} & Tandem &$15.1$ \\  \cline{2-3}	
   			& GMM-Mono & $17.1$ \\ \cline{2-3}	
   			& FH-Mono & $16.1$ \\  \cline{2-3}
   			& FS-FH-Mono & $18.8$ \\	\cline{1-3}	
   			Flat-start FH& Scratch & $19.0$ \\ \hline
   		\end{tabular}

   	\end{subtable}
   \end{table*}

  \begin{table*}[t]	
  	\setlength{\tabcolsep}{0.8em}\renewcommand{\arraystretch}{1.03}  
  	\centering \footnotesize \vspace{-0.3cm}
  	\caption{Results for our factored hybrid models compared to other phoneme based ASR systems trained on {\normalfont Switchboard 300h}.\ Separate results on {\normalfont Switchbaord~(SWB)} and {\normalfont CallHome~(CH)} using 4-gram LM, and WERs on of {\normalfont Hub'00} for both LSTM and 4-gram LM, are included.\ Label topology for transducer model is RNA and for the rest HMM.\ We also report number of epochs~(Eps.), where available.\ }
  	\vspace{-3mm}	
  	\label{tab:literature}
  	\begin{tabular}{|c|c|c|c|c|c|c|c|c|c|c|} 
  		\hline
  		\multirow{3}{*}{\textbf{\#}}&\multicolumn{3}{|c|}{\textbf{Model}} & \multirow{2}{*}{\textbf{State}}      & \multirow{3}{*}{\textbf{Context}} & \multicolumn{1}{c|}{\multirow{3}{*}{\textbf{Alignment}}}& \multicolumn{4}{c|}{\textbf{WER[\%]}} \\ \cline{2-4} \cline{8-11}
  		
  		& \multirow{2}{*}{\textbf{Name}}      & \multirow{2}{*}{\textbf{Training}}   & \multirow{2}{*}{\textbf{\#Ep.}}   & \multirow{2}{*}{\textbf{Tying}}                         && &\multicolumn{3}{c|}{\textbf{4-GRAM}}&\textbf{LSTM}  \\ \cline{8-11}
  		
  	   &&&&&&& \textbf{SWB}& \textbf{CH}&\multicolumn{2}{c|}{\textbf{Hub'00}}\\ \hline
  	    
  	    1&\multirow{2}{*}{Hybrid}              & \multirow{5}{*}{CE} &                                20 & \multirow{2}{*}{CART}                                   & Triphone &\multirow{4}{*}{Tandem} &$9.5$&$20.1$ & $13.9$ & $12.7$\\ \cline{1-1}\cline{4-4} \cline{6-6} \cline{8-11}  
  	    2&&&24&&\multirow{2}{*}{Diphone}& & $10.2$&$19.8$&$15.0$& \multirow{2}{*}{-}\\ \cline{1-2}\cline{4-5}\cline{8-10}  
  	    3&\multirow{3}{*}{Factored}&\multirow{3}{*}{with}&60&\multirow{5}{*}{No}&& & $9.6$&$18.5$&$14.1$&\\\cline{1-1}\cline{4-4} \cline{6-6}\cline{8-11}  
  	    4&\multirow{4}{*}{Hybrid}&\multirow{3}{*}{Viterbi}&\multirow{2}{*}{87}&&\multirow{6}{*}{Triphone}&&$\mathbf{9.2}$ &$17.7$& $13.5$&  $12.7$ \\ \cline{7-11} \cline{1-1} 
  	    5&\multirow{5}{*}{(this work)}&&&&&GMM-Mono& $10.2$&$19.4$&$14.8$&$13.8$\\\cline{7-11}  \cline{1-1}\cline{4-4}
  	    6&&&60&&&FH-Di& $9.8$&$18.6$&$14.2$ &$13.2$\\\cline{4-4}\cline{7-11} \cline{1-1} 
  	    7&&\multirow{5}{*}{Full-sum}&86&&&FS-FH-Mono&$10.7$&$19.2$&$15.0$& $13.9$\\ \cline{3-4}\cline{7-11}  \cline{1-1}
  	    8&&&60&&& \multirow{2}{*}{Scratch}&$12.1$&$25.9$&$19.0$& \multirow{4}{*}{-}\\\cline{1-2}\cline{4-5} \cline{8-10}
  	    9&\multicolumn{1}{|r|}{Hybrid\cite{zeyerbeckctc}} & & \multirow{2}{*}{-} & \multirow{2}{*}{CART} && & \multicolumn{2}{c|}{-} & $19.5$& \\  \cline{1-2} \cline{7-10} 
  	    10&\multicolumn{1}{|r|}{CTC \cite{patrick}}&&  &&&Hybrid  & $11.9$&$23.0$&$17.4$&\\ \cline{1-10}
  	    11&\multicolumn{1}{|r|}{Hybrid \cite{povey}} & LF-MMI& 4 &\multirow{3}{*}{No}&\multirow{2}{*}{Diphone} &Scratch &$9.8$&$19.3$& - &  \\ \cline{1-3}\cline{7-10}\cline{4-4}
  	    12&\multicolumn{1}{|r|}{Transducer \cite{wei}}&\multirow{2}{*}{CE} &50&&&CTC &$\mathbf{9.2}$&$\mathbf{17.6}$&$\mathbf{13.4}$& \\ \cline{1-2 }\cline{6-11} \cline{4-4}
  	    13&\multicolumn{1}{|r|}{Attention \cite{mohammad}}& &33&&Monophone&Scratch &\multicolumn{3}{c|}{-}& $14.4$\\ \hline
		
  	\end{tabular}   	
  	\vspace{-6mm}
  \end{table*}

\vspace{-0.3cm}
\section{Experimental results}
\vspace{-0.1cm}
 All models are trained and evaluated on 300h Switchboard-1 Release 2 (LDC97S62)~\cite{godfrey1992switchboard} and Hub'00 data (LDC2002S09), respectively, with the aid of RETURNN and RASR toolkits~\cite{zeyer2018returnn,wiesler2014rasr}.\ The Forward-Backward algorithm for computation of the soft alignments in the flat-start model, i.\ e., contexts and center state occupancies, makes use of a modified version of the CUDA based implementation presented in \cite{zeyerbeckctc}.\ 
All experiments share the same front-end architecture based on a Bi-LSTM encoder comprising 6 forward and backward layers of size 500 with $10\%$ dropout probability~\cite{srivastava2014dropout}.\ The input speech signal to the encoder is represented by 40-dimensional Gammatone Filterbank features~\cite{schluter2007gammatone}, extracted from 25 milliseconds~(ms) analysis frames with 10ms shift.\ Concerning the CD architectures, the one-hot encoding of the left and/or right phonemes and the center phoneme states are projected by using linear layers of dimension 10 and 30, respectively.\ Our proposed approach makes use of a state inventory consisting of three times number of phonemes plus the single-state silence entity.\ For the standard hybrid, a set of 9001 CART labels are considered.\ All models using Viterbi and fixed path share the same set of training hyper-parameters and are trained with the frame-wise CE criterion.\ During CE training, the sequences are divided into chunks of length $128$ with $50\%$ overlap.\ The chunking reduces overfitting as the model has access to a smaller context window during training and speeds up training.\ The full-sum training does not allow for chunking.\ The parameters for the flat-start model trained with full-sum~(FS) differ, as described below.\ All models are trained by using an Adam optimizer with Nesterov momentum~\cite{dozat2016incorporating}.\ We use Newbob learning rate~(LR) scheduling to reduce the initial LR of (CE: $1e^{-3}$, FS: $5e^{-4}$), with a decay factor of (CE: $0.9$ , FS: $\sqrt{0.8}$) based on (CE: frame error rate, FS: posterior scores) to a minumum value of (CE: $2e^{-5}$, FS:$1e^{-6}$).\ For further regularization, we use $L_2$ weight decay with a scale of $0.01$, gradient noise\cite{neelakantan2015adding} with a variance of (CE: $0.1$, FS: $0.3$) and focal loss factor of $2.0$~\cite{lin2017focal}.\ We normalize the state posteriors with the prior during full-sum training and during recognition of all models.\ During training, the prior estimation is done on-the-fly with exponential decaying average.\ The AM and prior scales for full-sum training starts with $0.01$ and $0.1$, respectively and grows at each epoch.\ We set maximum values of $0.3$ for AM scale, together with $0.3$ and $0.4$ for left and right contexts priors, and $0.7$ for center state phoneme prior.\ In \cite{zeyerbeckctc}, it is argued that if during full-sum training we do not apply state priors, the alignment will be dominated by silence.\ In our case, by using separate priors for each factor, we observed that the alignment contains very little silence compared to alignments from both GMM monophone and tandem systems.\ For recognition, priors are estimated by averaging over the output activations of the network using a subset of the training set.\ We utilize both 4-gram and LSTM language models for decoding~\cite{4gram,sundermeyer2012lstm,lstm}.\ 
 
An overview of different alignments and their description can be found in \cref{tab:align}.\ The GMM-Mono alignment derives from a monophone GMM system initialized with a linear alignment.\ For alignments taken from our factored hybrid models, we first train the monophone model of \cref{fig:monofact} with Viterbi and using GMM-Mono.\ By applying forced-alignment~(F-Align), we obtain the FH-Mono alignment.\ We then initialize a diphone model with the parameters of this trained model and continue training with GMM-Mono.\ A further F-align step using this model provides FH-Di.\ In different conducted experiments, we observed that independently from the alignment used for Viterbi training, the multi-stage phonetic training, i.\ e., initializing an n-phone model with the parameters of an (n-1)-phone model and training by resetting LR, leads always to improvement.\ This is shown for instance in \cref{tab:align}, where factored diphone achieves $9.9\%$ relative improvement compared to factored monophone, and in \cref{tab:literature} where triphone trained with Tandem~(row 4) has $4.9\%$ relative improvement compared to diphone~(row 3).\ A systematic comparison between monophone models trained with Viterbi using different alignments is reported in \cref{tab:mono}.\ We observed that realignment in an earlier phonetic stage is more useful.\ In the case of the monophone model it improves the performance from $17.1\%$ to $16.1\%$.\ Furthermore, given a flat-start model trained for 30 epochs, we can obtain a slightly better result by doing F-align and Viterbi training compared to full-sum.\ This is the difference between the last two models of \cref{tab:mono}.\ In \cref{tab:literature}, we compare our models with other phoneme-based approaches in the literature.\ Our flat-start factored hybrid by resetting LR and longer full-sum training~(row 8), reaches slightly better performance than the standard hybrid~(row $9$).\ We also F-aligned our flat-start model and carried out a 3-stage phonetic training with Viterbi.\ The resulting model~(row $7$) reached $15.0\%$, which is only $7.3\%$ relative worse compared to standard hybrid with state-tying trained with Viterbi.\ This degradation is limited to relative $8.4\%$ on SWB part of HUB'00 compared to flat-start hybrid with Lattice-Free MMI~(row $11$).\ However, this gap is increased to relative $10.6\%$ on HUB'00 compared to phoneme-based transducer~(row $12$) which uses a CTC alignment for Viterbi training.\ Differently to our flat-start factored hybrid, CTC does not use prior and transition probabilities during training.\ Regarding the factored hybrid models trained with Viterbi, it is possible to see that our triphone trained with GMM-Mono loses only $6.0\%$ relative compared to standard hybrid.\ This difference is reduced to $2.1\%$ relative when using FH-Di which requires one realignment at diphone stage.\ Finally, The factored triphone trained with Tandem alignment is $2.8\%$ relative better than standard hybrid trained with Viterbi and the same alignment.\ 
 \vspace{-0.35cm}
\section{Conclusions}
\vspace{-0.2cm}
We presented a flat-start factored hybrid model using the complete set of triphone states with similar performance to standard hybrid approach trained with full-sum training and state-tying.\ We also showed that our context-dependent factored hybrid using alignments from less powerful models, and trained with frame-wise cross-entropy using Viterbi, has only around $2\%$ relative degradation compared to standard hybrid using a tandem based alignment.\ Our proposed modeling framework considerably simplifies the standard pipeline in the hybrid approach.\ At the current state of the work we applied a simplifying assumption to our model trained with full-sum.\ We leave the introduction of the full triphone dependencies for training of our flat-start factored hybrid as a future work.\

\vspace{-0.35cm}
\section{Acknowledgements}
\vspace{-0.2cm}
This project has received funding from the European Research Council (ERC) under the European Union’s Horizon 2020 research and innovation programme (grant agreement n\textsuperscript{o}~694537, project "SEQCLAS"). The work reflects only the authors' views and the European Research Council Executive Agency (ERCEA) is not responsible for any use that may be made of the information it contains.\ The work was partly funded by the Google Faculty Research Award for "Label Context Modeling in Automatic Speech Recognition".\ Authors thank Wilfried Michel for insightful comments.

\bibliographystyle{IEEEtran}

\bibliography{mybib}

\end{document}